\documentclass[pdflatex,sn-mathphys-num,iicol]{sn-jnl}

\usepackage{booktabs}
\usepackage{multirow}
\usepackage{mathtools}
\usepackage{xcolor}
\usepackage{gensymb}
\usepackage{subfig}
\usepackage{siunitx}
\usepackage{chemformula}
\usepackage{amssymb}
\usepackage{lineno}

\def\u{$^{238}$U}
\def\th{$^{232}$Th}

\def\k{$^{40}$K}
\def\kqu{$^{41}$K}
\def\ka{$^{42}$K}
\def\na{$^{24}$Na}

\def\g{$\gamma$}
\def\b-{$\beta^-$}
\def\ppq{\SI{E-15}{g/g}}
\def\be{\begin{equation}}
\def\ee{\end{equation}}

\begin{document}

\title{Ultra-trace analysis of \k\ in organic liquid scintillators}

\author*[a]{A.~Barresi}\email{andrea.barresi@unimib.it}
\author[a]{D.~Chiesa}
\author[b]{D.~Merli}
\author*[a]{M.~Nastasi}\email{massimiliano.nastasi@unimib.it}
\author[c]{S.~Nisi}
\author[a,c]{E.~Previtali}
\author[a]{M.~Sisti}
\author[a]{M.~Borghesi}
\author[d]{A.~Cammi}
\author[a]{C.~Coletta}
\author[a]{G.~Ferrante}
\author[d]{L.~Loi}
\author[e]{G.~Andronico}
\author[f]{V.~Antonelli}
\author[f]{D.~Basilico}
\author[f]{M.~Beretta}
\author[g]{A.~Bergnoli}
\author[f]{A.~Brigatti}
\author[g]{R.~Brugnera}
\author[e]{R.~Bruno}
\author[h]{A.~Budano}
\author[f]{B.~Caccianiga}
\author[f]{A.~Caslini}
\author[g]{V.~Cerrone}
\author[e]{R.~Caruso}
\author[i]{C.~Clementi}
\author[g]{L.V.~D'Auria}
\author[g]{S.~Dusini}
\author[h]{A.~Fabbri}
\author[j]{G.~Felici}
\author[g]{A.~Garfagnini}
\author[f]{M.G.~Giammarchi}
\author[e]{N.~Giudice}
\author[g]{A.~Gavrikov}
\author[g]{M.~Grassi}
\author[e]{N.~Guardone}
\author[f]{F.~Houria}
\author[h]{A.~Islam}
\author[f]{C.~Landini}
\author[g]{L.~Lastrucci}
\author[g]{I.~Lippi}
\author[f]{P.~Lombardi}
\author[k,l]{F.~Mantovani}
\author[h]{S.M.~Mari}
\author[j]{A.~Martini}
\author[f]{L.~Miramonti}
\author[k,l]{M.~Montuschi}
\author[h]{D.~Orestano}
\author[i]{F.~Ortica}
\author[j]{A.~Paoloni}
\author[f]{L.~Pelicci}
\author[f]{E.~Percalli}
\author[h]{F.~Petrucci}
\author[f]{G.~Ranucci}
\author[f]{A.C.~Re}
\author[k,l]{B.~Ricci}
\author[i]{A.~Romani}
\author[g]{A.~Serafini}
\author[g]{C.~Sirignano}
\author[g]{L.~Stanco}
\author[h]{E.~Stanescu Farilla}
\author[k,l]{V.~Strati}
\author[f]{M.D.C~Torri}
\author[e]{C.~Tuve'}
\author[h]{C.~Venettacci}
\author[e]{G.~Verde}
\author[j]{L.~Votano}

\affil[a]{\orgname{INFN Sezione di Milano Bicocca e Dipartimento di Fisica, Università di Milano Bicocca}, \orgaddress{\city{Milano}, \country{Italy}}}
\affil[b]{\orgname{Department of Chemistry, University of Pavia}, \orgaddress{\city{Pavia},  \country{Italy}}}
\affil[c]{\orgname{INFN, Laboratori Nazionali del Gran Sasso}, \orgaddress{\city{Assergi}, \country{Italy}}}
\affil[d]{\orgname{INFN Sezione di Milano Bicocca e Dipartimento di Energetica, Politecnico di Milano}, \orgaddress{\city{Milano}, \country{Italy}}}
\affil[e]{\orgname{INFN Sezione di Catania e Università di Catania, Dipartimento di Fisica e Astronomia}, \orgaddress{\city{Catania}, \country{Italy}}}
\affil[f]{\orgname{INFN Sezione di Milano e Università degli Studi di Milano, Dipartimento di Fisica}, \orgaddress{\city{Milano}, \country{Italy}}}
\affil[g]{\orgname{INFN Sezione di Padova e Università di Padova, Dipartimento di Fisica e Astronomia}, \orgaddress{\city{Padova}, \country{Italy}}}
\affil[h]{\orgname{INFN Sezione di Roma Tre e Università degli Studi Roma Tre, Dipartimento di Matematica e Fisica}, \orgaddress{\city{Roma}, \country{Italy}}}
\affil[i]{\orgname{INFN Sezione di Perugia e Università degli Studi di Perugia, Dipartimento di Chimica, Biologia e Biotecnologie}, \orgaddress{\city{Perugia}, \country{Italy}}}
\affil[j]{\orgname{INFN, Laboratori Nazionali di Frascati}, \orgaddress{\city{Frascati}, \country{Italy}}}
\affil[k]{\orgname{INFN, Sezione di Ferrara}, \orgaddress{\city{Ferrara}, \country{Italy}}}
\affil[l]{\orgname{Università degli Studi di Ferrara, Dipartimento di Fisica e Scienze della Terra}, \orgaddress{\city{Ferrara}, \country{Italy}}}

\abstract{
Rare-event searches require exceptionally low background levels, motivating the development of increasingly sophisticated screening methods to push sensitivity limits. Liquid scintillators are particularly attractive detector media due to their intrinsic radiopurity and the ability to scale to large target masses. In this work, we present a screening strategy capable of measuring ultra-trace concentrations of \k\ with sensitivities below \SI{E-15}{g/g}. The method combines neutron activation analysis with a dedicated radiochemical treatment, followed by low-background HPGe gamma spectroscopy. Using this approach, we achieved a minimum detectable concentration of \SI{2.9E-16}{g/g} for \k, placing this technique among the most sensitive currently available.
}


\maketitle


\section{Introduction}
\label{S:Introduction}
Rare-event searches, particularly in neutrino physics, rely on large-scale detectors operating with extremely low background rates (see, e.g., \cite{juno-bkg,borexino-nature,sno+,gerda,cupid-mo,cuore-bkg,legend-1000} for a non-exhaustive list). In this context, background suppression has become one of the primary challenges, making the meticulous selection and validation of detector materials essential for minimizing spurious signals. As a result, laboratory screening techniques must achieve unprecedented sensitivity to certify materials at the levels demanded by modern experiments.

Among the available analytical methods, neutron activation analysis (NAA) stands out as a highly effective technique for detecting stable or long-lived nuclides such as \u, \th, and \k\ \cite{NAA-2024}. Its strength lies in converting the target nuclides into short-lived isotopes with high specific activity that can be readily quantified via gamma spectroscopy. When combined with low-background HPGe detectors, NAA typically achieves sensitivities at the \qty{E-12}{g/g} level for \u\ and \th, and of some \qty{E-15}{g/g} for \k\ in the absence of interfering activation products \cite{NAA-2024}. However, despite its excellent performance, this sensitivity is still insufficient for the newest rare-event experiments, which require concentrations below \SI{E-16}{g/g}.

In this work, we introduce a methodology designed to screen liquid scintillator (LS) matrices with sensitivities surpassing the \ppq\ threshold for \k. The technique integrates neutron activation analysis with dedicated radiochemical processing and high-efficiency HPGe gamma spectroscopy, enabling the analysis of irradiated samples under low-background conditions. The method is applied to a liquid scintillator based on linear alkyl benzene (LAB), adopted by the Jiangmen Underground Neutrino Observatory (JUNO) \cite{juno-ppnp}, though the procedure can be adapted to other scintillators used in future experiments. JUNO imposes stringent requirements on radiopurity \cite{juno-bkg}, and optical performance \cite{refractiveIndex-juno,fluorescence-juno}, which motivate the development of highly sensitive screening techniques such as the one presented in this work.

The paper is organized as follows. Section~\ref{S:NAAMeasurementStrategy} outlines the principles of neutron activation analysis and details our measurement strategy. Section~\ref{S:RadioChemProcedureAndSampleMeasurement} describes each step of the radiochemical process and the sample measurement. Section~\ref{S:RecoveryEfficiency} reports the tests performed to assess the recovery efficiency of the radiochemical treatment. Section~\ref{S:SampleAndSensitivity} presents the analysis of a liquid scintillator sample together with the estimated sensitivity.

\section{Neutron activation and measurement strategy}
\label{S:NAAMeasurementStrategy}
NAA is a powerful technique that provides exceptional sensitivity for the detection of stable or long-lived nuclides, such as \k\ \cite{NAA-2024}. 

The direct activation product of \k\ is \kqu, which is a stable nuclide. For this reason, potassium analysis by neutron activation relies on the \kqu\ isotope naturally present in potassium with an isotopic abundance of \qty{6.73}{\%}. Upon neutron capture, \kqu\ produces \ka, according to the reaction:
\begin{align*}
    &^{41}\text{K} \xrightarrow{(n,\gamma)} \, ^{42}\text{K} \xrightarrow[12.36\text{ h}]{\beta^-} \, ^{42}\text{Ca}
\end{align*}
The activity of \ka\ is then quantified by gamma-ray spectroscopy through the measurement of the \qty{1524.6}{keV} $\gamma$ line, emitted with a probability of \qty{18.08}{\%}, allowing one to infer the concentrations of \kqu\ and subsequently \k\ by assuming natural isotopic composition. Further details on the fundamentals of neutron activation analysis can be found in \cite{NAA-2024,UTh-2026}.

Direct NAA applied to a liquid scintillator (LS) sample already provides excellent sensitivity to \k, typically at the level of some \qty{E-15}{g/g} when no interfering activation products are present. To further improve the achievable sensitivity, we developed a dedicated measurement procedure specifically optimized for ultra-trace potassium determination.

The core idea is to apply a radiochemical treatment to the LS after irradiation, with the dual purpose of concentrating the potassium contained in the sample and suppressing the presence of interfering nuclides. This step is essential because several elements—most notably sodium—produce activation products that emit high-energy gamma rays, whose Compton continuum overlaps with the region of interest of activated potassium and significantly degrades the measurement sensitivity. Since no commercially available resin enables a selective separation of potassium from sodium and other chemically similar elements, a gravimetric (precipitation-based) approach was adopted.

Crucially, all radiochemical operations are performed after neutron irradiation to avoid external contamination. Potassium is both abundant in nature and a key biological element, making even trace accidental introduction a significant risk.

The proposed methodology enhances the detectability of potassium by increasing its effective concentration in the final sample while markedly suppressing competing backgrounds, thus enabling sensitivities well beyond those obtainable with direct NAA alone.

\begin{figure*}
    \centering
    \includegraphics[width=0.9\textwidth]{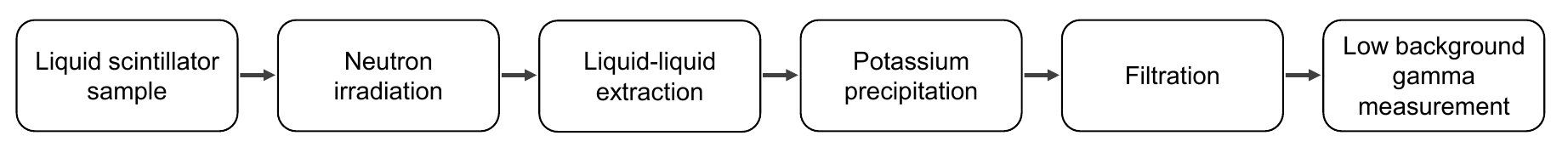}
    \caption{Block diagram of the measurement procedure for potassium.}
    \label{fig:BlockDiagram}
\end{figure*}
The block diagram in Figure~\ref{fig:BlockDiagram} illustrates the conceptual sequence of the procedure.

    \subsection{Labware cleaning protocol}
    \label{S:LabwareCleaning}
    The maximum allowable concentration of \k\ in liquid scintillators, at the \qty{E-16}{g/g} level required by modern rare-event experiments\cite{juno-bkg}, is many orders of magnitude lower than the common abundance of potassium in the environment. Achieving such ultra-low sensitivities makes contamination control essential, particularly for the materials and containers used to collect, store, and irradiate the samples.

In the methodology developed for this work, with radiochemical treatments performed after neutron irradiation, the strictest cleanliness requirements apply to the containers used during sample collection and handling, as well as to the irradiation vials, which must not introduce any additional potassium before activation. Extensive testing allowed us to identify the most suitable materials for these operations. We employ PFA (perfluoroalkoxy) for both the bottles and the irradiation vials, which are subjected to a dedicated cleaning protocol. Further details on material selection and cleaning procedures can be found in \cite{UTh-2026}.

    \subsection{Neutron irradiation and HPGe pre-screening}
    \label{S:NeutronIrradiation}
    The neutron activation of the samples is carried out at the TRIGA Mark II research reactor of the University of Pavia, located roughly \SI{40}{km} from the University of Milano-Bicocca. This pool-type facility, cooled and partially moderated by light water, operates at a maximum nominal power of \qty{250}{kW}. The reactor core—cylindrical in shape, about \qty{46}{cm} in diameter and \qty{36}{cm} in height—is surrounded by a \qty{30}{cm}-thick radial graphite reflector, with additional axial reflectors placed above and below the fuel region. Several irradiation positions are available, including the Central Thimble at the core center and the Lazy Susan carousel embedded in the radial reflector. Further details about the reactor configuration and its irradiation facilities can be found in Refs.~\cite{AbsoluteFlux,BayesianSpectrum,FluxDistribution}.

The Lazy Susan system is particularly convenient thanks to its 40 irradiation channels, which provide substantial flexibility in the number of samples that can be activated in a single run. Each channel is cylindrical, with an internal diameter of approximately \qty{32}{mm}. For our irradiations, we employ custom polyethylene irradiation containers with an inner diameter of \qty{25}{mm} and a usable height of about \qty{100}{mm}, designed to accommodate the PFA vials used for sample handling. Typical neutron fluxes at the Pavia TRIGA reactor are on the order of \qty{1.7E13}{n/cm^2/s} in the Central Thimble and \qty{2.2E12}{n/cm^2/s} in the Lazy Susan carousel~\cite{BayesianSpectrum}. The reactor typically operates with daily irradiation periods of approximately six hours at full power, and our NAA campaigns are organized around this schedule.

Since the sensitivity achievable with NAA strongly depends on the mass of the irradiated sample, it is advantageous to activate the largest possible amount of material. For the methodology developed for ultra-trace \k\ measurements, we typically irradiate a total LS mass of about \SI{100}{g}. Due to the geometric constraints of the irradiation channels, this amount is divided into five PFA vials with a nominal volume of \qty{30}{mL}, each placed in a separate channel. An additional channel is used to irradiate a natural potassium standard solution, required for the comparative determination of the potassium concentration. Furthermore, AlCo wire samples with a cobalt concentration of \qty{0.5}{\% (g/g)} and a typical mass of \qty{0.01}{g} are inserted into each irradiation container to monitor and correct for flux non-uniformity among different channels. More details about this method can be found in \cite{NAA-2024}.

The transfer of the LS from the \qty{500}{mL} PFA bottle into the irradiation vials is performed in a Class 10000 (ISO7) cleanroom, under a laminar flow hood that reduces the local cleanliness level to Class 1000 (ISO6). This procedure minimizes the risk of introducing potassium contaminants during the pouring process.

Since the radiochemical treatment cannot be performed on the same day of irradiation, at the end of the neutron irradiation, the five LS vials are emptied into five polyethylene (PE) bottles and measured overnight with HPGe detectors for a preliminary screening. Owing to the limited sample mass and to the presence of activation products from interfering nuclides, the sensitivity of this screening is on the order of \qty{1e-14}{g/g}.

Only the samples that show no evidence of anomalous potassium content after about \qty{10}{h} of measurement are subsequently combined and subjected to radiochemical processing. This selection is justified by the observation that both “clean” and “contaminated” samples could sometimes occur within the same batch: in these circumstances, the measured contaminations cannot be intrinsic to the liquid scintillator itself, but must arise from accidental contamination during handling or sample preparation.

\section{Radiochemical procedure and sample measurement}
\label{S:RadioChemProcedureAndSampleMeasurement}

    \subsection{Radiochemical procedure}
    \label{S:RadioChemProcedure}
    The developed radiochemical procedure consists of two consecutive stages: liquid–liquid extraction followed by selective precipitation. For a \SI{100}{mL} LS sample, the procedure comprises the following steps:
\begin{enumerate}
\item Liquid–liquid extraction of the LS performed three times at room temperature, each with \SI{10}{mL} of \SI{0.1}{M} acetic acid and \qty{0.5}{mg} of potassium carrier in the form of a \SI{5}{mg/mL} KCl water solution 
\item Addition of \qty{1.0}{mL} of \qty{0.1}{M} sodium tetraphenylborate (Na-TPB);
\item Waiting for complete precipitation of potassium tetraphenylborate (K-TPB);
\item Filtration of the precipitate on a \qty{0.45}{\micro\m} nylon syringe filter;
\item Washing of the filter with \SI{2}{mL} of cold K-TPB–saturated water.
\end{enumerate}

The liquid–liquid extraction step is used to transfer potassium from the organic scintillator matrix into an aqueous phase. This process exploits the different affinities of the solute for two immiscible solvents: an apolar organic phase (the LAB-based LS) and an aqueous phase consisting of an acetic acid solution. Potassium migrates into the aqueous phase, where it forms water-soluble salts. Acetic acid acts both as a buffer and as a competing agent for any long-chain carboxylic acid potassium complexes present in the organic phase, promoting their transfer into the aqueous phase through protonation and a mass-action-driven equilibrium shift. 

Besides enabling a phase transfer, the extraction step also provides a pre-concentration of the analyte, since the total volume of the extracting solution is smaller than that of the original sample. This pre-concentration is essential for the subsequent precipitation stage, which cannot be performed directly in an organic medium.

When two immiscible solvents are brought into contact, solutes redistribute between the phases until equilibrium is reached. This equilibrium is described by the distribution coefficient $\kappa$, defined as the ratio of the solute concentration in the organic phase to that in the aqueous phase. For a given analyte–solvent system, $\kappa$ depends on parameters such as temperature and pH, but not on the relative phase volumes. As a result, extraction efficiency is increased by performing multiple successive extractions with small volumes of fresh solution rather than a single extraction with a larger volume.

At the same time, the extracting solution must remain sufficiently voluminous to ensure effective contact between the two phases. In practice, a total aqueous-to-organic volume ratio of approximately 1:3 is adopted. Accordingly, \qty{30}{mL} of extracting solution is used for \qty{100}{mL} of LS, divided into three aliquots of \qty{10}{mL} each. Each extraction step lasts approximately \qty{20}{minutes}, during which the mixture is vigorously stirred with a magnetic stirrer to promote efficient phase interaction.

Following extraction, a selective precipitation is performed to isolate \ka\ from other activated nuclides, concentrate it, and maximize the detection efficiency in the final HPGe measurement. As an alkali metal, potassium forms only a limited number of insoluble compounds; among these, potassium tetraphenylborate (K-TPB) is particularly suitable due to its low solubility (\SI{56}{mg/L}) and its widespread use in analytical potassium determinations.

To enhance the overall extraction yield and ensure efficient precipitation, \qty{1.5}{mg} of natural (non-irradiated) potassium is added as a carrier during each extraction. After combining the aqueous extracts, \qty{1.0}{mL} of \qty{0.1}{M} Na-TPB solution is added to selectively precipitate potassium as K-TPB, while sodium and other interfering elements remain in solution. After approximately \qty{20}{minutes}, complete crystal formation is achieved, and the precipitate is collected on a \qty{0.45}{\micro\m} nylon syringe filter.

\begin{figure}
    \centering
    \includegraphics[width=0.45\textwidth]{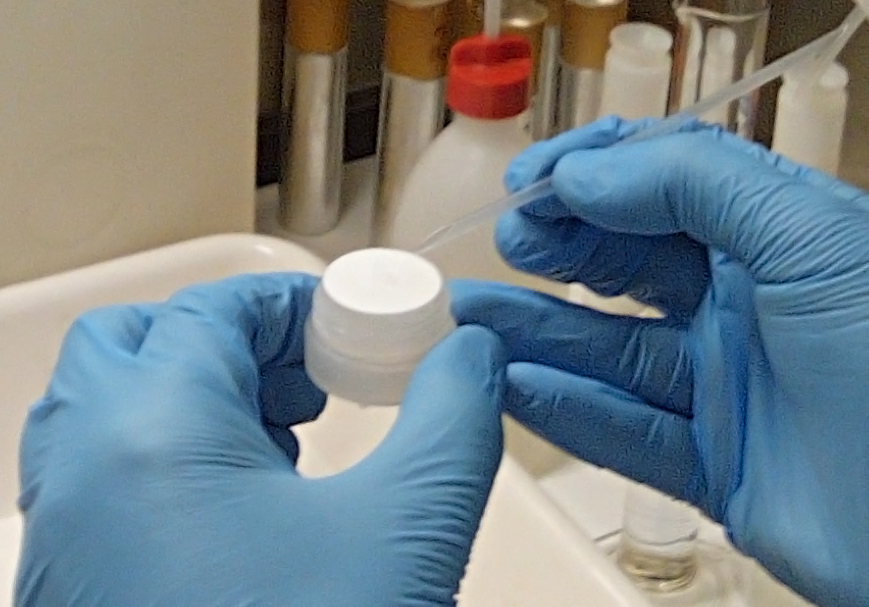}
    \caption{Nylon filter on its support used to collect the K-TPB precipitate.}
    \label{fig:Filter}
\end{figure}

The presence of the potassium carrier ensures the formation of a sufficient amount of K-TPB with crystals large enough to be retained by the filter. The filter is subsequently washed with \SI{2}{mL} of cold K-TPB–saturated water to remove residual sodium and other contaminants while minimizing analyte losses through the common-ion effect. Finally, the filter containing the potassium precipitate is transferred into small polyethylene vials and analyzed using a HPGe detector.

It is worth noting that alternative conditions were also evaluated. In particular, experiments conducted using a \ch{NaNO3} extraction solution at \SI{0.1}{M} in combination with a thallium carrier resulted in significantly lower recovery efficiency. Moreover, strong acids such as \ch{HNO3} cannot be employed for precipitation with TPB, as they induce its decomposition.
    
    \subsection{HPGe gamma spectroscopy}
    \label{S:GammaSpectroscopy}
    To determine the activity of \ka\ in the filter obtained from the radiochemical procedure, we employed a well-type HPGe detector with a resolution of \qty{2.3}{keV} at \qty{1332}{keV} and a crystal volume of \qty{350}{cm^3}. This detector features a germanium crystal with a blind hole in which the sample is inserted, achieving an almost $4\pi$ geometry and thus maximizing the absolute counting efficiency.

The sample is measured for approximately \SI{18}{h}, corresponding to about 1.5 half-lives of \ka. At the end of the acquisition, the spectrum is examined for the presence of the \SI{1525}{keV} peak. According to the method of Currie~\cite{Currie:1968}, two scenarios are possible: if the number of counts in the peak exceeds the critical level, the potassium concentration in the sample is calculated; otherwise, an upper limit on the concentration is derived based on the background in the region of interest.

\section{Recovery efficiency determination}
\label{S:RecoveryEfficiency}
Both the precipitation and filtration processes described in Section~\ref{S:RadioChemProcedure} are characterized by an efficiency. Knowledge of this efficiency is crucial, as it directly affects the final sensitivity of the measurements. In this section, we describe the methodology adopted to determine the efficiency of these treatments and present the corresponding results.

To assess the process efficiency, LS samples were spiked with a known quantity of activated natural potassium, and the recovered amount was measured after completing the full procedure. A certified solution of natural potassium from Inorganic Ventures was irradiated simultaneously, but separately from the LS sample. A known amount of the irradiated potassium was then introduced after irradiation to ensure precise control over the amount added to the sample. Since this solution is too concentrated for direct use and is immiscible with the organic LS matrix, a stepwise dilution protocol was implemented:
\begin{itemize}
    \item Dilution 1:100: mix \SI{0.10}{mL} of the K certified solution (\SI{1000}{\micro g/g} in 1\% \ch{HNO3}) with \SI{9.9}{mL} of 1\% acidified water to obtain \SI{10}{mL} of a \SI{10}{\micro g/g} K solution;
    \item Irradiation of \SI{1.0}{mL} of the \SI{10}{\micro g/g} K solution;
    \item Dilution 1:100: dilute \SI{0.10}{mL} of the irradiated solution in \SI{9.9}{mL} of isopropyl alcohol to obtain a \SI{100}{\nano g/g} solution.
\end{itemize}

The final dilution employs isopropyl alcohol, which is miscible with the non-polar LS. A volume of \SI{0.10}{mL} of this solution contains \SI{1.0E-8}{g} of natural potassium, corresponding to \SI{1.17E-12}{g} of \k\ when accounting for its isotopic abundance (0.0117\%). This amount was added to \SI{20}{g} of irradiated LS, yielding a contamination level of approximately \SI{6E-14}{g/g} of \k. The mixture was homogenized using a magnetic stirrer for several minutes to ensure uniform dispersion of the nuclides throughout the LS matrix.

Twelve LS samples prepared in this manner were analyzed across four efficiency evaluation campaigns. To verify the initial contamination level, aliquots of the spiking solution were independently measured with HPGe detectors, confirming that the introduced potassium mass was consistent with the expected value in all cases. The LS samples were then processed according to steps 1--5 outlined in Section~\ref{S:RadioChemProcedure}, and the final filters were analyzed by \g\ spectroscopy to quantify the recovered potassium.

\begin{figure*}
    \centering
    \includegraphics[width=0.8\textwidth]{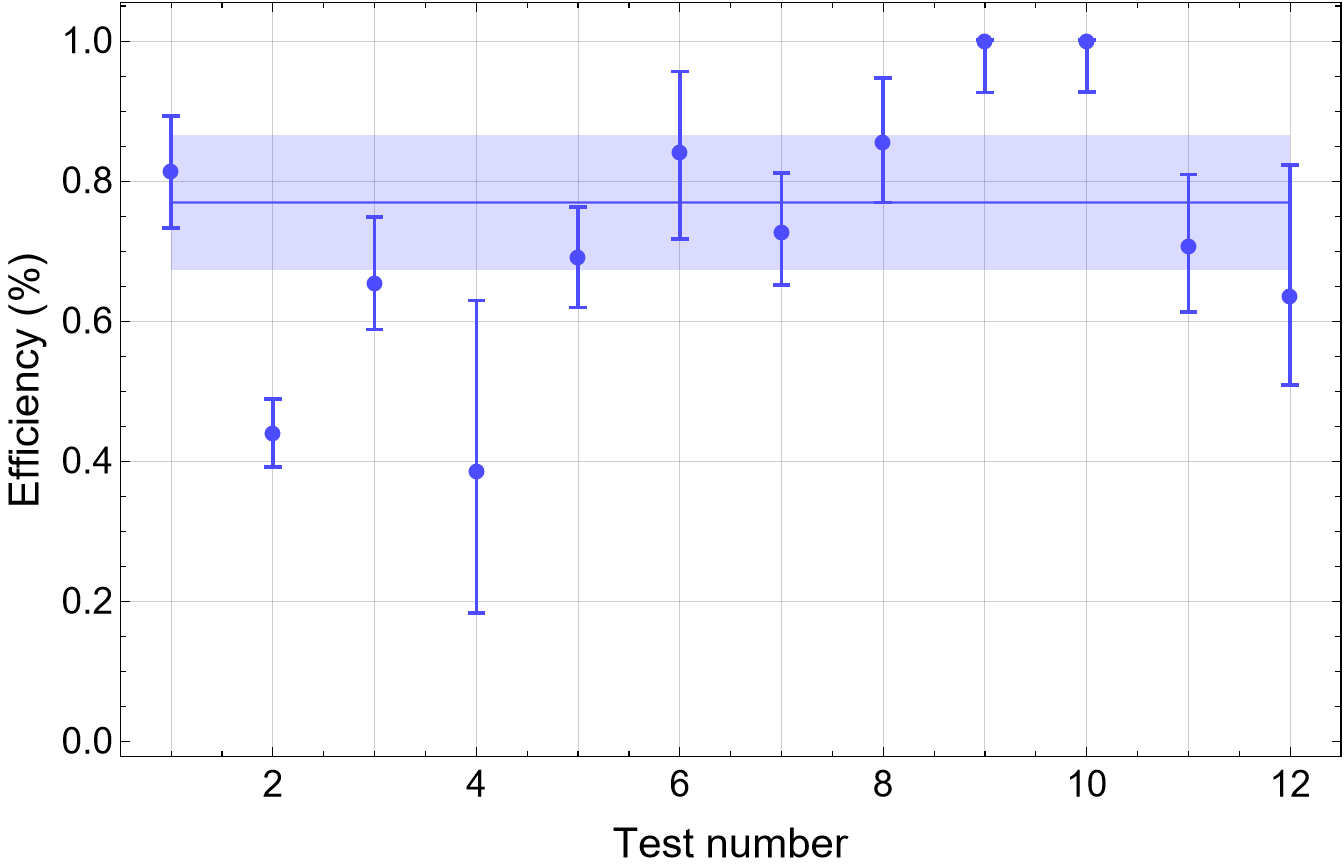}
    \caption{Recovery efficiency of the complete potassium procedure. The line represents the mean value, and the band shows the corresponding uncertainty calculated as the standard deviation of the measurements.}
    \label{fig:Efficiency}
\end{figure*}

The efficiencies are computed by exploiting a Markov chain Monte Carlo approach to correctly limit the efficiencies and their errors between zero and one. The results are reported in Table~\ref{tab:CompleteEfficiencies} and Figure~\ref{fig:Efficiency}. Potassium recovery shows high efficiency and good reproducibility. The uncertainty on the efficiency was calculated as the standard deviation of the individual measurements, yielding a final efficiency of \SI{77(10)}{\%}.
The lower recovery observed in samples 2 and 4 is attributed to losses during the filtration step, caused by filter clogging and partial leakage of the suspension from the syringe.

\begin{table}[t]
    \centering
    \begin{tabular}{cc}
        \toprule
        Test number & K efficiency (\%) \\
        \midrule
        1 & \num{81(8)} \\
        2 & \num{44(5)} \\
        3 & \num{65(8)} \\
        4 & \num{38(22)} \\
        5 & \num{69(7)} \\
        6 & \num{84(12)} \\
        7 & \num{73(8)} \\
        8 & \num{85(8)} \\
        9 & $100^{+0}_{-7}$ \\
        10 & $100^{+0}_{-7}$ \\
        11 & \num{71(1)} \\
        12 & $64^{+19}_{-13}$ \\
        Mean & \num{77(10)} \\
        \bottomrule
    \end{tabular}
    \captionof{table}{Results of the total efficiency of the radiochemical process for potassium. The last row reports the mean of the measurements, with the error computed as the standard deviation of the results.}
    \label{tab:CompleteEfficiencies}
\end{table}

\section{Sample measurement and sensitivity evaluation}
\label{S:SampleAndSensitivity}
A common approach to evaluate the sensitivity of a measurement technique is to apply it to a blank sample. For the procedure described here, preparing a blank sample is not feasible because the LS is directly irradiated without prior treatment, and no LS with a certified radiopurity in potassium below our sensitivity goal is available, and any other material would not accurately represent the behavior of the LS.

For this reason, we report a measurement of a JUNO liquid scintillator sample collected during the commissioning phase of the purification plants \cite{impianti}; since it was taken at this preliminary stage, the sample does not necessarily reflect the final radiopurity achieved in the detector. This measurement, however, reaches the highest sensitivity of the method and is therefore representative of its expected performance, allowing us to assess the impact of residual interfering nuclides.

\begin{figure*}
    \centering
    \includegraphics[width=0.95\textwidth]{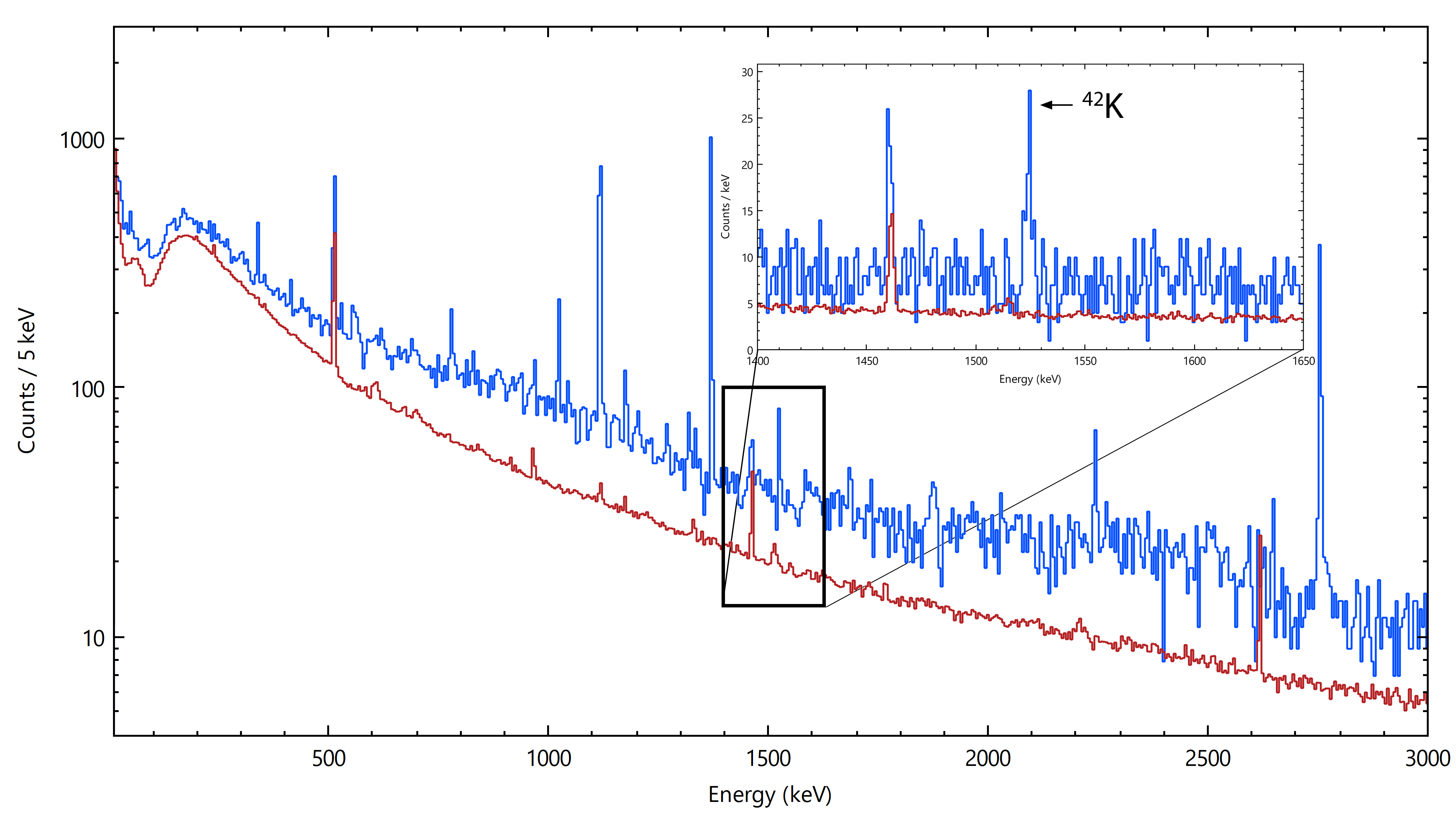}
    \caption{HPGe spectrum of the filter obtained after the radiochemical procedure applied to the irradiated LS sample (blue), superimposed to the normalized HPGe background spectrum (red). The inset shows a zoom around the region of interest at \SI{1525}{keV}, highlighting the \ka\ peak. In the spectrum are clearly visible other peaks due to residual contamination of $^{115}$Cd (\qty{336}{keV}), $^{65}$Zn (\qty{1115}{keV}), and $^{24}$Na (\qty{1368}{keV} and \qty{2754}{keV})}
    \label{fig:SampleSpectrum}
\end{figure*}

The LS sample was collected at the end of the purification line into a \SI{500}{mL} PFA bottle to minimize possible contamination. Due to the constraints of the irradiation channel size, as described in Section~\ref{S:NeutronIrradiation}, the LS was divided into five PFA vials and irradiated for \SI{6}{h} under a thermal neutron flux of approximately \SI{2.2e12}{n/cm^2/s}. After irradiation, the five samples were measured individually overnight using HPGe detectors to check for anomalous \ka\ concentrations that might have arisen from potential contamination during the transfer from the collection bottle to the irradiation vials. 
None of the five samples showed a \ka\ concentration above the preliminary measurement sensitivity (see Table \ref{tab:MeasurementResults}), so all samples were then combined and processed using the radiochemical procedure, yielding a total mass of measured LS of \SI{99.8}{g}. The resulting nylon filter was measured with a well-type HPGe detector for \SI{17.5}{h}, producing the spectrum shown in Figure~\ref{fig:SampleSpectrum}.

The main interfering nuclide is \na, whose intense gamma-ray emission at \SI{2754}{keV} (B.R.\ 99.9\%) increases the background in the region of interest and thus degrades the sensitivity to \ka. As described in Section~\ref{S:NAAMeasurementStrategy}, one of the objectives of the radiochemical treatment is to remove \na\ and suppress its contribution to the gamma spectrum.

\begin{table*}[t]
    \centering
    \begin{tabular}{ccccc}
        \toprule
        Sample & Sample mass & \k\ concentration & Na concentration & Na mass\\
        & (g) & (\SI{E-15}{g/g}) & (\SI{E-12}{g/g}) & (ng) \\
        \midrule
        1 & \num{20.7} & \num{<6.7} & \num{13.2(1.2)} & \num{0.27(0.02)} \\
        2 & \num{19.4} & \num{<11} & \num{50.7(3.4)} & \num{0.98(0.06)} \\
        3 & \num{19.8} & \num{<6.4} & \num{108(6)} & \num{2.1(0.1)} \\
        4 & \num{19.6} & \num{<7.7} & \num{538(28)} & \num{10.5(0.5)} \\
        5 & \num{20.3} & \num{<8.2} & \num{95(5)} & \num{1.92(0.11)} \\
        Sum 1-5 & \num{99.8} & - & - & \num{15.9(0.6)} \\
        Filter & - & \num{0.65(0.16)} & - & \num{0.271(0.015)} \\
        \midrule
        Efficiency & & & & \SI{98.3(0.1)}{\%} \\
        \bottomrule
    \end{tabular}
    \captionof{table}{Summary of \k\ and Na concentrations measured in five aliquots of the same LS sample and in the filter after radiochemical treatment.}
    \label{tab:MeasurementResults}
\end{table*}

By comparing the total sodium content in the LS samples after irradiation with the amount remaining in the final filter, the removal efficiency can be determined.
The mass of Na was determined by using as a reference standard the AlCo wires flux monitors, considering the effective cross section of Na and Co of \qty{0.29(0.03)}{barn} and \qty{21.40(0.07)}{barn}, respectively. 
In Table~\ref{tab:MeasurementResults}, the measured concentrations of Na in each of the five LS samples before the radiochemical procedure and in the filter after potassium precipitation, together with the corresponding masses, are reported. The amount of Na before treatment is obtained by summing the Na mass across all LS samples and comparing it to the amount observed in the filter.
The resulting removal efficiency is \SI{98.3(0.1)}{\%}, demonstrating that the radiochemical procedure effectively suppresses the dominant interfering nuclide and substantially improves the sensitivity to detect \ka.

During the measurement, a peak at \SI{1525}{keV}, corresponding to the gamma emission of \ka, was observed, indicating the presence of potassium. A fit of the peak yielded the number of counts used to calculate the \k\ concentration in the LS, resulting in \SI{6.5(1.6)e-16}{g/g}.

To estimate the sensitivity of the measurement technique, the mean background near the region of interest was evaluated, giving a background index of \SI{6.9(0.2)}{counts/keV}. This value is only slightly higher than the intrinsic background index of the HPGe detector, measured to be \SI{3.74(0.04)}{counts/keV} in the same energy interval and normalized for measurement time, as also visible in Figure~\ref{fig:SampleSpectrum}.
Achieving a residual background level so close to the detector’s intrinsic background was made possible by the high $^24\text{Na}$ removal efficiency of \SI{98.3(0.1)}{\%}, confirming that the radiochemical procedure effectively suppresses the dominant source of spectral interference in the ROI and allows the sensitivity to be limited primarily by the detector performance.

The number of counts in the ROI in the absence of signal was calculated by multiplying the background index by the detector resolution (FWHM) and by a factor of 1.25 \cite{DEGEER2004151}, which defines an interval containing \qty{84}{\%} of the signal. This value was then used to compute the minimum detectable concentration, using the detection limit computed according to Equation \ref{eq:LD} \cite{Currie:1968}, yielding \SI{2.9e-16}{g/g}.
\begin{equation}
\label{eq:LD}
    L_D = 2.71+3.29\sqrt{\mu_B}
\end{equation}

\section{Conclusion}
\label{S:Conclusion}
In this paper, we presented a radiochemical procedure combined with neutron activation analysis to measure ultra-trace amounts of \k\ in liquid scintillators. By transferring potassium from the organic LS to an aqueous phase, selectively precipitating it as K-TPB, and measuring the resulting filter with a well-type HPGe detector, the methodology effectively concentrates potassium while suppressing interfering nuclides.

The efficiency of the complete radiochemical process was evaluated using spiked LS samples, yielding a mean potassium recovery of \SI{77(10)}{\%}. Applying the full procedure to a commissioning JUNO LS sample, we determined a \k\ concentration of \SI{6.5(1.6)e-16}{g/g}. The estimated sensitivity, derived from the background near the region of interest, corresponds to a minimum detectable concentration of \SI{2.9e-16}{g/g}.

These results demonstrate that the developed approach surpasses the \SI{E-15}{g/g} threshold, providing one of the most sensitive screening techniques for \k\ currently available. The methodology can be readily adapted to other liquid scintillators, and further improvements could be achieved by optimizing the precipitation and filtration steps or increasing the sample mass during irradiation, potentially enhancing both sensitivity and reproducibility.

\bibliography{Bibliography}

\end{document}